\documentclass[twoside,prd,showpacs,nofootinbib,preprintnumbers,twocolumn]{revtex4-1}

\usepackage{amssymb,amsmath,bm,natbib}
\usepackage{color}
\usepackage{slashed}
\usepackage{graphics}
\usepackage{graphicx}
\usepackage[utf8]{inputenc}
\usepackage{upgreek}
\usepackage[caption=false]{subfig}
\usepackage{hyperref}
\usepackage{url}
\usepackage{dsfont}
\usepackage{float} 
\usepackage{cancel}
\usepackage{units}
\usepackage{blindtext}
\usepackage[dvips]{feynmp}
\def\ring#1{{\mathaccent'27 #1}}

\begin{document}
\preprint{ZU-TH 40/22, PSI-PR-22-25}
	\title{Impact of Lorentz Violation on Anomalous Magnetic Moments of Charged Leptons	}
	
\author{Andreas Crivellin}
\email{andreas.crivellin@cern.ch}
\affiliation{Paul Scherrer Institut, CH--5232 Villigen PSI, Switzerland}
\affiliation{Physik-Institut, Universit\"at Z\"urich, Winterthurerstra\ss{}e 190, CH--8057 Z\"urich, Switzerland}

\author{Fiona Kirk}
\email{fiona.kirk@psi.ch}
\affiliation{Paul Scherrer Institut, CH--5232 Villigen PSI, Switzerland}
\affiliation{Physik-Institut, Universit\"at Z\"urich, Winterthurerstra\ss{}e 190, CH--8057 Z\"urich, Switzerland}
		
\author{Marco Schreck}
\email{marco.schreck@ufma.br}
\affiliation{Departamento de F\'{i}sica, Universidade Federal do Maranh\~{a}o \\
Campus Universit\'{a}rio do Bacanga, S\~{a}o Lu\'{i}s (MA), 65085-580, Brazil}
			
\begin{abstract}
We address the question whether a violation of Lorentz symmetry can explain the tension between the measurement and the Standard-Model prediction of the anomalous magnetic moment of the muon ($(g-2)_{\upmu}$) and whether it can significantly impact the one of the electron ($(g-2)_{\mathrm{e}}$). While anisotropic Lorentz-violating effects are, in general, expected to produce sidereal oscillations in observables, isotropic Lorentz violation (LV) in the charged-lepton sector could feed into $(g-2)_{\mathrm{e},\upmu}$. However, we find that this type of Lorentz violation, parametrised via a dim-4 field operator of the Standard-Model Extension (SME), is already strongly constrained by the absence of vacuum \v{C}erenkov radiation and photon decay. In particular, the observations of very-high-energetic astrophysical photons at LHAASO and of high-energetic electrons (muons) by the LHC (IceCube) place the most stringent two-sided bounds on the relevant SME coefficients $\ring{c}^{(\mathrm{e})}$ $(\ring{c}^{(\upmu)})$. Therefore, any explanation of the tension in $(g-2)_{\upmu}$ via isotropic Lorentz violation of the minimal spin-degenerate SME is excluded, and the possible size of its impact on $(g-2)_{\mathrm{e}}$ is very limited.
\end{abstract}



\maketitle

\newpage
\begin{fmffile}{diagrams}
\section{Introduction}

Anomalous magnetic moments of charged leptons pose excellent tests of quantum field theory, dating back to Schwinger's famous prediction $a_\ell\equiv (g-2)_\ell/2=\alpha/(2\pi)\simeq 1.16\times 10^{-3}$~\cite{Schwinger:1948iu} and its later experimental confirmation~\cite{Kusch:1948mvb} for electrons. Currently, the comparisons between direct measurements~\cite{Hanneke:2008tm} and the Standard-Model (SM) predictions yield for the electron
\begin{subequations}
\label{eq:g-2_e}
\begin{align}
\label{eq:g-2_e_val_1}
\Delta a_{\mathrm{e}}[\text{Cs}]&\equiv a_{\mathrm{e}}^\text{exp}-a_{\mathrm{e}}^\text{SM}[\text{Cs}]=-0.88(28)(23)[36]\times 10^{-12}\,, \\
\label{eq:g-2_e_val_2}
\Delta a_{\mathrm{e}}[\text{Rb}]&\equiv a_{\mathrm{e}}^\text{exp}-a_{\mathrm{e}}^\text{SM}[\text{Rb}]=+0.48(28)(9)[30]\times 10^{-12}\,,
\end{align}
\end{subequations}
depending on whether the fine-structure constant $\alpha$ is taken from Cs~\cite{Parker:2018vye} or Rb~\cite{Morel:2020dww} atom interferometry. The errors refer to $a_{\mathrm{e}}^\text{exp}$, $\alpha$, and the total one, respectively. For the muon, the experimental average~\cite{Bennett:2006fi,Abi:2021gix,Albahri:2021ixb,Albahri:2021kmg,Albahri:2021mtf} differs from the SM prediction of the corresponding white paper~\cite{Muong-2:2021ojo}:\footnote{This result is based on Refs.~\cite{Aoyama:2012wk,Aoyama:2019ryr,Czarnecki:2002nt,Gnendiger:2013pva,Davier:2017zfy,Keshavarzi:2018mgv,Colangelo:2018mtw,Hoferichter:2019gzf,Davier:2019can,Keshavarzi:2019abf,Kurz:2014wya,Melnikov:2003xd,Masjuan:2017tvw,Colangelo:2017fiz,Hoferichter:2018kwz,Gerardin:2019vio,Bijnens:2019ghy,Colangelo:2019uex,Blum:2019ugy,Colangelo:2014qya}. Note that the recent lattice results~\cite{Borsanyi:2020mff,Alexandrou:2022amy,Ce:2022kxy} for hadronic vacuum polarization (HVP), which would render the SM prediction of $a_{\upmu}$ compatible with experiment at the $2\,\sigma$ level, are not included. However, these lattice results are in tension with HVP determined from $\mathrm{e^+e^-}\to$ hadrons data~\cite{Davier:2017zfy,Keshavarzi:2018mgv,Colangelo:2018mtw,Hoferichter:2019gzf,Davier:2019can,Keshavarzi:2019abf}. Furthermore, HVP enters the global electroweak fit~\cite{Passera:2008jk}, whose (indirect) determination is below the lattice values~\cite{Ce:2022eix,Colangelo:2022vok}, increasing the tension within the electroweak fit~\cite{Crivellin:2020zul,Keshavarzi:2020bfy}.}
\begin{equation}
\Delta a_{\upmu}\equiv a_{\upmu}^\text{exp}-a_{\upmu}^\text{SM}=2.51(59)\times 10^{-9}\,,
\label{eq:g-2_mu_vals}
\end{equation}
which constitutes a $4.2\,\sigma$ tension.

Many extensions of the SM by new particles have been proposed to explain the anomalous magnetic moment of the muon (see Ref.~\cite{Athron:2021iuf} for a recent review). We refer to, most prominently, the MSSM (see, e.g.~Refs.~\cite{Lopez:1993vi,Chattopadhyay:1995ae,Dedes:2001fv,Stockinger:2006zn}), models with generic new scalars and fermions~\cite{Stockinger:1900zz,Giudice:2012ms,Dermisek:2013gta,Falkowski:2013jya,Altmannshofer:2016oaq,Kowalska:2017iqv,Crivellin:2018qmi,Arnan:2019uhr,Crivellin:2021rbq}, $\mathrm{Z}^\prime$ bosons~\cite{He:1990pn,He:1991qd,Altmannshofer:2016brv,Buras:2021btx}, leptoquarks~\cite{Djouadi:1989md, Chakraverty:2001yg,Cheung:2001ip,Bauer:2015knc,ColuccioLeskow:2016dox,Dorsner:2019itg,Crivellin:2020tsz,Crivellin:2020mjs,Fajfer:2021cxa,Greljo:2021xmg} or additional Higgs doublets~\cite{Cao:2009as,Broggio:2014mna,Wang:2014sda,Ilisie:2015tra,Abe:2015oca,Crivellin:2015hha}, etc. However, there is also the possibility of modifying quantum field theory itself at the fundamental level instead of adding new particles. The physics responsible for such modifications is usually associated with the Planck scale and its impact on $(g-2)_{\ell}$ was studied, e.g.~in Ref.~\cite{Cohen:1998zx}. They found, though, that these effects were not capable of explaining the muon anomaly~\cite{Cohen:2021zzr}. In addition, an explanation of the tension in $(g-2)_\upmu$ by nonlocal QED was discussed in Ref.~\cite{Capolupo:2022awe}.

Alternatively, Lorentz-violating physics, which may occur at the Planck scale, could be able to account for $\Delta a_{\upmu}$. Here, the aforementioned SME \cite{Colladay:1996iz,Colladay:1998fq} provides a comprehensive effective-field theory framework to parametrise possible LV in all particle sectors of the SM. This approach allows us to assess the implications of LV at energy scales currently accessible to experiments. The SME incorporates operators involving SM fields whose Lorentz indices are suitably contracted with background fields. The latter give rise to preferred spacetime directions and the potential size of LV is encoded in the Wilson coefficients, which are known as the (controlling) coefficients in the SME literature. Therefore, the SME constitutes a modified quantum field theory whose theoretical aspects have been extensively investigated~\cite{Reyes:2008rj,Reyes:2008jn,Reyes:2010pv,Klinkhamer:2010zs,Klinkhamer:2011ez,Schreck:2011ai,Lopez-Sarrion:2012eri,Reyes:2013nca,Schreck:2013gma,Schreck:2013kja,Maniatis:2014xja,Schreck:2014qka,Reyes:2014wna,Reis:2016hzu,Nascimento:2017ebc}.

The literature comprises only few studies of LV effects on $(g-2)_{\ell}$ beyond the SME, e.g.~Refs.~\cite{Hayakawa:1999zf,Mouchrek-Santos:2018vrn}.
In an SME setting, $(g-2)_{\ell}$ has been investigated with LV in Dirac fermions \cite{Bluhm:1997qb,Bluhm:1998qf,Bluhm:1998zh,Bluhm:1999dx,Chen:2000zh,Muong-2:2001xzf,Muong-2:2007ofc,Kostelecky:2013rta,Gomes:2014kaa,Aghababaei:2017bei,Quinn:2019ppv,Lin:2021cst}, photons~\cite{Carone:2006tx,Moyotl:2013via}, Yukawa couplings~\cite{Ahuatzi-Avendano:2020ppm}, and the Higgs field~\cite{Castulo:2022gpn}. The focus of Refs.~\cite{Bluhm:1997qb,Bluhm:1998qf,Bluhm:1998zh,Bluhm:1999dx,Chen:2000zh,Muong-2:2001xzf,Muong-2:2007ofc,Kostelecky:2013rta,Gomes:2014kaa,Quinn:2019ppv,Lin:2021cst} is on modified spin precession governed by the spin-nondegenerate coefficients $b,d,H$, and $g$. Unless for very specific choices of coefficients, this approach leads to time-dependent effects in the lab, which cannot account for the discrepancies in Eqs.~\eqref{eq:g-2_e}, \eqref{eq:g-2_mu_vals}.
The coefficients known as $c$, which are CPT-conserving and describe spin-independent effects, are expected to provide a possible explanation of the tension in $(g-2)_{\upmu}$. After all, their isotropic part does not lead to sidereal oscillations. Until now, only the authors of Ref.~\cite{Aghababaei:2017bei} performed such a study. However, our tree-level result disagrees with theirs, which also has consequences on the physics.

\section{Lorentz-violating QED}

The SME parametrises LV modifications of the SM in a model-independent way, preserving gauge invariance and, thus, the Ward identities.
Before electroweak symmetry breaking, there are four distinct types of controlling coefficients contracted with power-counting renormalizable operators in the SME lepton sector~\cite{Colladay:1996iz,Colladay:1998fq}. The coefficients $(a_L)^{\mu}$ and $(c_L)^{\mu\nu}$ couple to left-handed $\textit{SU}(2)$ doublets, whereas $(a_R)^{\mu}$ and $(c_R)^{\mu\nu}$ couple to right-handed $\textit{SU}(2)$ singlets. After electroweak symmetry breaking, four new types of coefficients emerge in the charged-lepton sector from linear combinations of $(a_{L,R})^{\mu}$ and $(c_{L,R})^{\mu\nu}$. The first two are denoted as $(a^{\ell})^{\mu},(c^{\ell})^{\mu\nu}$ and are accompanied by a parity-even Dirac bilinear, while $(b^{\ell})^{\mu},(d^{\ell})^{\mu\nu}$ come with a parity-odd Dirac bilinear~\cite{Crivellin:2020oov}.

These coefficients alter free-particle propagation of charged leptons, which even differs for the two possible spin projections in the presence of the $b$-type and $d$-type coefficients~\cite{Colladay:1996iz,Kostelecky:2013rta}. The extended free Dirac theory is minimally coupled to the photon sector, which gives rise to a modified quantum electrodynamics (QED). In our analysis we do not need to consider the $a$ coefficients, since they can be removed in QED by (individual) redefinitions of the charged-lepton fields~\cite{Colladay:1996iz,Colladay:1998fq}. The impact of the spin-nondegenerate $b$ and $d$ coefficients on $(g-2)_{\ell}$ has been the subject matter of a fair number of papers \cite{Bluhm:1997qb,Bluhm:1998qf,Bluhm:1998zh,Bluhm:1999dx,Chen:2000zh,Muong-2:2001xzf,Muong-2:2007ofc,Kostelecky:2013rta,Gomes:2014kaa,Quinn:2019ppv,Lin:2021cst}.

Here, we focus on the spin-degenerate $c$ coefficients, which affect leptons and antileptons in the same way \cite{Kostelecky:2008bfz} and enter the anomalous magnetic moment both via their effect on the free propagation of leptons and via the modified QED vertex. Since the $c$ coefficients relate to coefficients in photons that are denoted as $k_F$ \cite{Altschul:2006zz}, we consider the following Lagrange density:
\begin{align}
\label{eq:modified-qed}
\mathcal{L}&=\frac{1}{2}\overline{\psi}\left[\mathrm{i}(\gamma^{\mu}+c^{\nu\mu}\gamma_{\nu})D_{\mu}-m\right]\psi+\text{h.c.} \notag \\
&\phantom{{}={}}-\frac{1}{4}(\eta_{\mu\varrho}\eta_{\nu\sigma}+(k_F)_{\mu\nu\varrho\sigma})F^{\mu\nu}F^{\varrho\sigma}\,.
\end{align}
All fields are defined in Minkowski spacetime with the metric tensor $\eta_{\mu\nu}$ of signature $(+,-,-,-)$. Furthermore, $\psi$ is a Dirac field standing for a single charged lepton, $\overline{\psi}=\psi^{\dagger}\gamma^0$ the Dirac-conjugate field, $m$ the fermion mass and $\gamma^{\mu}$ are the standard Dirac matrices satisfying the Clifford algebra $\{\gamma^{\mu},\gamma^{\nu}\}=2\eta^{\mu\nu}$. Furthermore, $D_{\mu}=\partial_{\mu}+\mathrm{i}qA_{\mu}$ denotes the gauge-covariant derivative for a particle of charge $q$ (with $q=-e$, $e>0$ for electrons) and $F_{\mu\nu}=\partial_{\mu}A_{\nu}-\partial_{\nu}A_{\mu}$ is the electromagnetic field strength tensor with the \textit{U}(1) gauge field $A_{\mu}$.

For convenience, we dropped the index $\ell$ for $(c^{\ell})^{\mu\nu}$, which refers to the charged-lepton flavour. Apart from $c$ coupling to Dirac fermions, the modified QED of Eq.~\eqref{eq:modified-qed} involves yet another background field known as $k_F$ and affecting photons. The flavour-universal part of $c$ and a subset of the $k_F$ coefficients can inherently be mapped onto each other~\cite{Altschul:2006zz}. Hence, it makes sense to perform a combined study of their effects on $(g-2)_{\ell}$, which provides an important cross-check of the result to be obtained.

Since the Lagrange density is Hermitian by definition, $c^{\mu\nu}$ and $(k_F)_{\mu\nu\varrho\sigma}$ must be real. Furthermore, the antisymmetric part of $c^{\mu\nu}$ can be removed by a field redefinition~\cite{Kostelecky:2003fs}:
\begin{equation}
\psi(x)\mapsto \psi'(x)=\left(1+\frac{\mathrm{i}}{4}c^{\mu\nu}\sigma_{\mu\nu}\right)\psi(x)\,,
\end{equation}
with $\sigma^{\mu\nu}\equiv(\mathrm{i}/2)[\gamma^{\mu},\gamma^{\nu}]$. Also, the trace of $c^{\mu\nu}$ is taken to be zero in a canonically normalized kinetic term. The background field $k_F$ transforms as a four-tensor of rank 4 under coordinate transformations, satisfying
\begin{equation}
(k_F)_{\mu\nu\kappa\lambda}=(k_F)_{\kappa\lambda\mu\nu}=-(k_F)_{\nu\mu\kappa\lambda}=-(k_F)_{\mu\nu\lambda\kappa}\,,
\end{equation}
as well as a Bianchi-type identity $(k_F)_{\kappa(\lambda\mu\nu)}=0$. The latter holds upon summation over cyclic permutations of a triple of its indices. Furthermore, the double trace of $k_F$ can be removed by another field redefinition, such that $(k_F)^{\mu\nu}_{\phantom{\mu\nu}\mu\nu}=0$ is assumed. The generic constraints on LV~\cite{Kostelecky:2008bfz} strongly suggest perturbative modifications, i.e.~$|c^{\mu\nu}|\ll 1$ and $|(k_F)_{\mu\nu\varrho\sigma}|\ll 1$. Hence, it is sufficient to work at first order in the coefficients.

\section{Modified anomalous magnetic moment}

\begin{figure}
\centering
\subfloat[]{\label{fig:tree-level-amplitude-first}\includegraphics{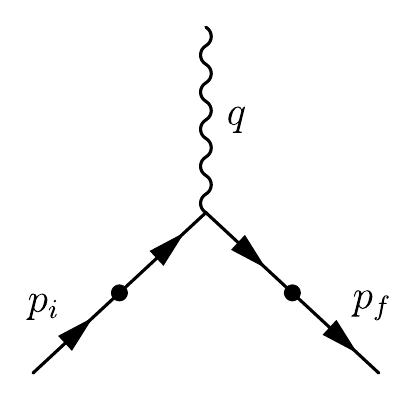}}\subfloat[]{\label{fig:tree-level-amplitude-second}\includegraphics{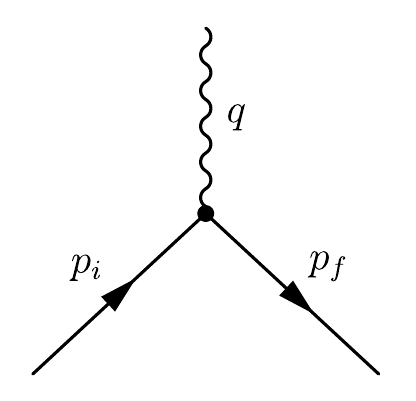}}
\caption{Contributions responsible for $a_{\ell}=(g-2)_\ell/2$ based on Eq.~\eqref{eq:modified-qed}. We use the conventions $p_i=(P+q)/2$ and $p_f=(P-q)/2$ for the coming and outgoing fermion momentum, respectively, and $q\equiv p_i-p_f$ is the outgoing-photon momentum. The bullets indicate LV insertions, i.e.~the modification of the propagation of the charged lepton and the coupling to the photon, respectively.}
\label{fig:tree-level-amplitude}
\end{figure}

Without LV, the tree-level amplitude can be rewritten by applying the Gordon identity
\begin{align}
\label{eq:amplitude-tree-level-standard}
\mathrm{i}\mathcal{M}^{(0)}_{ss'}&=-\overline{u}^{(s)}\mathrm{i}e\gamma^{\mu}u^{(s')}A_{\mu}(q) \notag \\
&=-\frac{\mathrm{i}e}{2m}\overline{u}^{(s)}\left(P^{\mu}A_{\mu}(q)-\frac{1}{2}\sigma_{\mu\nu}F^{\mu\nu}(q)\right)u^{(s')}\,,
\end{align}
with the four-momenta $P^{\mu}$ and $q^{\mu}$, as defined in Fig.~\ref{fig:tree-level-amplitude}, and where $\overline{u}^{(s)}=\overline{u}^{(s)}(p_f,m)$, $u^{(s')}=u^{(s')}(p_i,m)$ are understood. To perform an analysis of $(g-2)$ physics, $A_{\mu}$ is replaced by a static vector potential $\overline{A}_{\mu}$, describing the limit of a slowly varying external magnetic field with components $B^i$. Then, $\overline{A}_0=0$ and $-\mathrm{i}\varepsilon^{ijk}q_i\overline{A}^j(q)=B^k$, whereby an expansion around $q^{\mu}=0$ is permissible. Since our sole interest is in the tensorial contribution, which arises at linear order in $q^{\mu}$, we can simply keep the constant (mass-dependent as opposed to momentum-dependent) contribution from the spinors such that
\begin{equation}
\overline{u}^{(s)}\sigma^{0i}u^{(s')}=0\,,\quad \overline{u}^{(s)}\sigma^{ij}u^{(s')}=\varepsilon^{ijk}\xi^{(s)\dagger}\sigma^k\xi^{(s')}\,,
\end{equation}
where $\xi^{(s)}$ are two-component spinors with $\xi^{(s)\dagger}\xi^{(s')}=\delta^{ss'}$. We employ the spin operator for particle states, $\mathbf{S}=\boldsymbol{\sigma}/2$, with the set of Pauli matrices $\boldsymbol{\sigma}=(\sigma^x,\sigma^y,\sigma^z)$. The spin quantization axis is set to point along the magnetic field where we choose the coordinate system such that $\mathbf{B}=B\hat{\mathbf{z}}$ \cite{Muong-2:2021ojo}. Then, the part of the amplitude that the anomalous magnetic moment can be deduced from is
\begin{equation}
\label{eq:amplitude-tree-level-standard-spin-conserving}
\mathcal{M}^{(0)}_{ss}\supset \frac{e}{4m}\overline{u}^{(s)}\sigma_{\mu\nu}u^{(s)}F^{\mu\nu}(q)=-\frac{e}{m}BS^z\,,
\end{equation}
with the spin eigenvalues $S^z=\pm 1/2$ along the quantization axis. Comparing Eq.~\eqref{eq:amplitude-tree-level-standard-spin-conserving}
to the nonrelativistic potential $V(x)=-\boldsymbol{\mu}\cdot\mathbf{B}(x)$ with $\boldsymbol{\mu}=g_{\ell}e/(2m)S^z\hat{\mathbf{z}}$, we identify Dirac's famous prediction $g_{\ell}=2$, which holds at tree-level and when LV corrections are disregarded.

Our next step is to compute $a_{\ell}$ in the presence of $c$-type LV with $k_F=0$. The modified Dirac equation in momentum space based on Eq.~\eqref{eq:modified-qed} then reads
\begin{equation}
(\cancel{p}+c^{\mu\nu}\gamma_{\mu}p_{\nu}-m)u(p)=0\,,
\end{equation}
where $u(p)$ is a particle spinor. The modified Dirac equation gives rise to a modified Gordon identity
\begin{align}
\overline{u}(p_f)\gamma^{\mu}u(p_i)&=\frac{1}{2m}\overline{u}(p_f)\Big[(\eta^{\mu\lambda}+c^{\mu\lambda})P_{\lambda} \notag \\
&\phantom{{}={}}\hspace{0.9cm}-\mathrm{i}\sigma^{\mu}_{\phantom{\mu}\kappa}(\eta^{\kappa\lambda}+c^{\kappa\lambda})q_{\lambda}\Big]u(p_i)\,,
\end{align}
where we dropped the spin index, as $a_{\ell}$ follows from the spin-conserving contributions, anyhow.
The modification of the Gordon identity is illustrated by the Feynman diagram in Fig.~\ref{fig:tree-level-amplitude-first}. It is judicious to absorb all coefficients $c^{\mu\nu}$ that enter the propagators and the spinors into the momenta of the fermions:
\begin{equation}
\overline{p}^{\mu}\equiv p^{\mu}+c^{\mu\nu}p_{\nu}\,.
\end{equation}
In terms of the redefined momentum, the modified Dirac equation and Gordon identity take the same form as in the SM:
\begin{subequations}
\begin{align}
(\cancel{\overline{p}}-m)u(\overline{p})&=0\,, \\
\overline{u}(\overline{p}_f)\gamma^{\mu}u(\overline{p}_i)&=\frac{1}{2m}\overline{u}(\overline{p}_f)(\overline{P}^{\mu}-\mathrm{i}\sigma^{\mu\nu}\overline{q}_{\nu})u(\overline{p}_i)\,.
\end{align}
\end{subequations}
The complete tree-level amplitude also involves a modified fermion-photon coupling induced by the covariant derivative according to the diagram of Fig.~\ref{fig:tree-level-amplitude-second}. By defining $\Gamma^{\mu}\equiv\gamma^{\mu}+c^{\nu\mu}\gamma_{\nu}$, we write the complete tree-level amplitude for modified Dirac fermions in the form
\begin{subequations}
\begin{align}
\mathrm{i}\mathcal{M}_{\psi,ss'}^{(0)}&=-\overline{u}^{(s)}\mathrm{i}e\Gamma^{\mu}u^{(s')}A_{\mu}(q) \notag \\
&=\mathrm{i}(\mathcal{M}^{(0)}_{ss'}+\delta\mathcal{M}_{\psi,ss'}^{(0)})\,,
\end{align}
with the standard contribution given in Eq.~\eqref{eq:amplitude-tree-level-standard} and the modification
\begin{align}
\delta \mathcal{M}_{\psi,ss'}^{(0)}&=-\frac{e}{2m}\overline{u}^{(s)}\bigg[-\frac{1}{2}(c^{\mu\lambda}\sigma_{\lambda}^{\phantom{\lambda}\nu}-c^{\nu\lambda}\sigma_{\lambda}^{\phantom{\lambda}\mu})F_{\mu\nu}(q) \notag \\
&\phantom{{}={}}\hspace{1.6cm}{}+2c^{\mu\nu}P_{\nu}A_{\mu}(q)\bigg]u^{(s')}\,.
\end{align}
\end{subequations}
Applying the procedure described at the beginning of the current section, we find the following LV modification of the amplitude in Eq.~\eqref{eq:amplitude-tree-level-standard-spin-conserving}:
\begin{align}
\delta\mathcal{M}_{\psi,ss}^{(0)}&\supset \frac{e}{4m}\overline{u}^{(s)}(c^{\mu\lambda}\sigma_{\lambda}^{\phantom{\lambda}\nu}-c^{\nu\lambda}\sigma_{\lambda}^{\phantom{\lambda}\mu})u^{(s)}F_{\mu\nu}(q) \notag \\
&=\frac{e}{m}(c^{11}+c^{22})BS^z\,.
\end{align}
Finally, the tree-level contribution of the $c$ coefficients to the anomalous magnetic moment, at leading order in LV, reads
\begin{equation}
\label{eq:anomaly-modified-fermions}
a_{\psi}\equiv-\frac{m}{e}\frac{\delta\mathcal{M}_{\psi,ss}^{(0)}}{BS^z}=-(c^{11}+c^{22})\,.
\end{equation}
Thus, only a small subset of purely spacelike $c$ coefficients is capable of inducing an anomalous magnetic moment at tree-level. It is this combination of coefficients that introduces LV into the cyclotron frequency \cite{Bluhm:1997qb,Bluhm:1998qf,Bluhm:1998zh}, i.e. we interpret Eq.~\eqref{eq:anomaly-modified-fermions} as not arising from modified spin precession; cf. Refs.~\cite{Kostelecky:2013rta,Gomes:2014kaa}.

Parametrising LV in the photon sector by a specific subset of $k_F$ implies a result similar to that of Eq.~\eqref{eq:anomaly-modified-fermions}; cf.~App.~\ref{sec:modified-photons}. This outcome is expected, since an appropriate coordinate transformation can remove the $c$ coefficients from the fermion sector, whereby the $k_F$ coefficients get shifted accordingly~\cite{Altschul:2006zz}. This finding provides an independent crosscheck of Eq.~\eqref{eq:anomaly-modified-fermions}. Since LV in photons is tightly constrained by spectropolarimetry \cite{Friedman:2018ubt,Kislat:2018rsi,Friedman:2020bxa,Gerasimov:2021chj} as well as table-top experiments with optical cavities \cite{Michimura:2013kca,Nagel:2014aga}, we will discard $k_F$ in the following. Also, it is beyond the scope of this work to investigate sidereal variations of $(c^{\ell})^{\mu\nu}$. Hence, as of now, we will be working with the setting of an isotropic $c$-type background field:
\begin{align}
c^{\mu\nu} = \ring{c}\, \mathrm{diag}\left(1,\frac{1}{3},\frac{1}{3},\frac{1}{3}\right)^{\mu\nu}, \qquad \ring{c}\equiv c^{00}\,.
\end{align}
In this case, Eq.~\eqref{eq:anomaly-modified-fermions}
takes the form\footnote{This means that isotropic LV, contrary to the findings of Ref.~\cite{Aghababaei:2017bei}, leads to an effect in the anomalous magnetic moments of charged leptons.}
\begin{align}
    a_\psi = -\frac{2}{3}\ring{c}\,.
\end{align}
Confronting this result with Eq.~\eqref{eq:g-2_mu_vals}, the discrepancy between the experimental and the SM predictions for $a_{\upmu}$ could be explained by the presence of isotropic LV in muons with
\begin{equation}
\label{eq:value-c-muon-from-g-2}
    \ring{c}^{(\upmu)}=-3.8\times 10^{-9}\,,
\end{equation}
while for electrons
\begin{equation}
\label{eq:values-c-eletron-from-g-2}
    \ring{c}^{(\mathrm{e})}[\mathrm{Cs}]= 1.3\times 10^{-12}\,,\quad \ring{c}^{(\mathrm{e})}[\mathrm{Rb}]=-7.2\times 10^{-13}\,.
\end{equation}

\section{Vacuum \v{C}erenkov radiation and photon decay}

In this section we will assess whether the size of LV needed to explain $\Delta a_{\upmu}$ or leading to a relevant effect in $a_{\mathrm{e}}$ is compatible with constraints from other processes. Firstly, charged, massive particles with an energy above a specific threshold lose energy via vacuum \v{C}erenkov radiation~\cite{Beall:1970rw,Coleman:1997xq,Moore:2001bv,Lehnert:2004be,Lehnert:2004hq,Kaufhold:2005vj,Kaufhold:2007qd,Altschul:2006zz,Altschul:2007kr,Hohensee:2008xz,Klinkhamer:2008ky,Altschul:2014bba,Schober:2015rya,Diaz:2015hxa,Kostelecky:2015dpa,Colladay:2016rmy,Altschul:2016ces,Colladay:2016rsf,Colladay:2017auq,Schreck:2017isa,Altschul:2017xzx,Schreck:2017egi,Schreck:2018qlz}. In particular, for the isotropic controlling coefficient $\ring{c}$ this happens when the charged-lepton energy exceeds~\cite{Schreck:2017isa}
\begin{align}
\label{eq:threshold-energy-isotropic-c}
    E^\mathrm{th}_{\psi} = \frac{1}{2} \sqrt{\frac{3}{2}} \frac{m}{\sqrt{-\ring{c}}}\,.
\end{align}
The minus sign inside the square root indicates that this process is possible only for $\ring{c}<0$. Furthermore, the radiated-energy rate near the threshold is~\cite{Schreck:2017isa}\footnote{This result corresponds to the estimate on the right-hand side of Eq.~(13) in Ref.~\cite{Hohensee:2008xz} apart from the dimensionless prefactor. The latter can only be properly derived from a quantum field theoretic treatment of the process.}
\begin{align}
\label{eq:radiated-energy-rate}
\frac{\mathrm{d}W}{\mathrm{d}t}&\simeq \frac{64}{9}\sqrt{\frac{2}{3}}\alpha(-\ring{c})^{3/2}\frac{(E-E^{\mathrm{th}}_{\psi})^3}{m} \notag \\
&\simeq \frac{4}{3}\alpha m^2\left(\frac{E-E^{\mathrm{th}}_{\psi}}{E}\right)^3\,.
\end{align}
This result allows us to conclude that the process is very efficient with a typical radiation length of $\ll \unit[1]{m}$. If a propagating particle does not suffer energy losses, the particle energy must lie below the threshold of Eq.~\eqref{eq:threshold-energy-isotropic-c}, which, in turn, restricts the maximum size of $-\ring{c}$.
\begin{figure*}
    \centering
    \includegraphics[width=.9\linewidth]{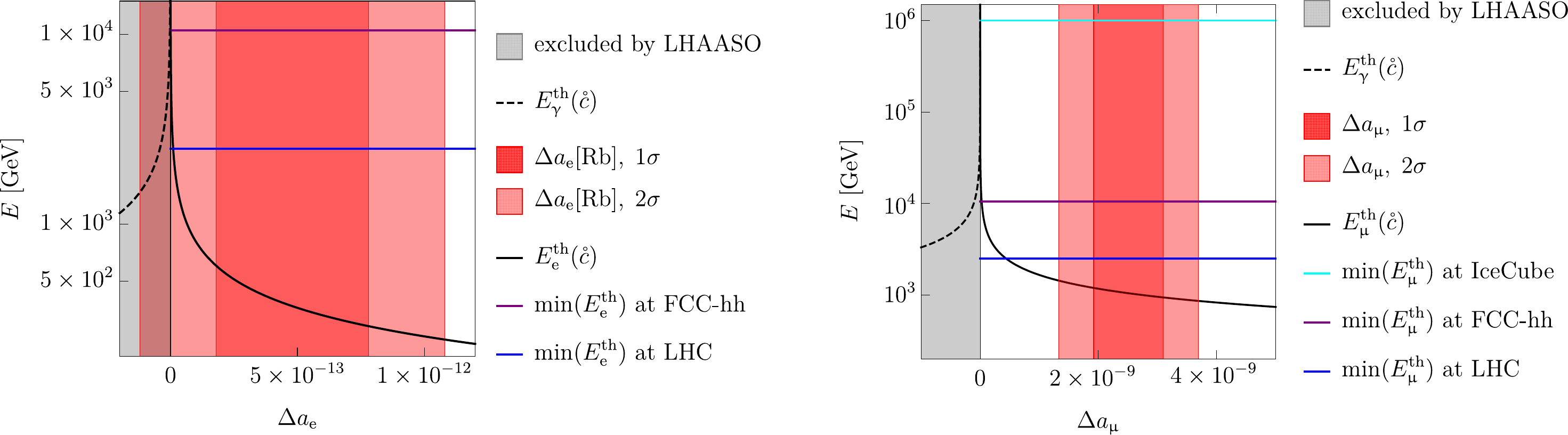}
    \caption{Left (right): Difference of the experimental average and the SM prediction of the anomalous magnetic moment of the electron (muon), $\Delta a_{\mathrm{e}}$ ($\Delta a_{\upmu}$), versus the charged-lepton energy. The threshold energy for photon decay, $E_{\upgamma}^\mathrm{th}$, in Eq.~\eqref{eq:threshold-energy-photon-decay-isotropic-c} is illustrated by a dashed, black curve and the regions of $\Delta a_{\mathrm{e}(\upmu)}$ excluded by photon decay are coloured in grey. We do not show $\Delta a_\mathrm{e}[\mathrm{Cs}]$, since the whole $2\sigma$-preferred region is excluded by LHAASO. The threshold energy for vacuum \v{C}erenkov radiation,  $E_{\mathrm{e}(\upmu)}^\mathrm{th}$, given in Eq.~\eqref{eq:threshold-energy-isotropic-c} is represented by a solid, black curve, whereas the lepton energies probed by IceCube, by the LHC, and by the proposed Future Circular Hadron Collider (FCC-hh) are indicated by horizontal lines. The regions to the right of the $\Delta a_{\mathrm{e}(\upmu)}$-value corresponding to the intersection of the curve $E_\mathrm{e}^\mathrm{th}$ ($E_\upmu^\mathrm{th}$) with the LHC (IceCube) line are excluded. We observe that the IceCube measurements exclude the values of $\ring{c}^{(\upmu)}$ that would be required to explain the tension in $\Delta a_{\upmu}$.}
    \label{fig:electron_muon_plot}
\end{figure*}

Therefore, constraints on $\ring{c}$ can be obtained from the maximum charged-lepton energies observed experimentally. The highest energy of electrons produced in a particle physics laboratory was measured by CMS at the LHC in the search for a charged lepton plus missing transverse momentum (see, e.g.~Fig.~4 of Ref.~\cite{CMS:2022yjm}). It amounts to around $\unit[2.5]{TeV}$, which implies that the threshold energy must be lower than this value. Hence, we infer the conservative lower bound
\begin{equation}
    \label{eq:constraint-c-electron-LHC}
    -1.6\times 10^{-14}<\ring{c}^{(\mathrm{e})}\,.
\end{equation}
Note that this LHC bound is more stringent than that obtained previously from sychrotron radiation at LEP~\cite{Hohensee:2008xz}.\footnote{The cosmic-electron energy spectrum measured by several experiments~\cite{AMS:2000kgm,Boezio:2001dtm,Grimani:2002yz,HESS:2008ibn,PAMELA:2011bbe,Fermi-LAT:2011baq}, with CALET~\cite{CALET:2017uxd} and AMS~\cite{AMS:2019iwo} providing the most recent analyses, could also be consulted to constrain LV via the absence of vacuum \v{C}erenkov radiation. Note that HESS detected a $\unit[3]{TeV}$ electron event \cite{HESS:2008ibn}. However, since their analysis involves simulations for disentangling proton and photon events, we will use the maximum electron energy measured at the LHC to constrain $\ring{c}^{(\mathrm{e})}$ conservatively from below. Anyhow, the potential constraint obtained from the $\unit[3]{TeV}$ data point would only be marginally better than that of Eq.~\eqref{eq:constraint-c-electron-LHC}.} In a survey of the atmospheric-muon spectrum, IceCube measured the highest-energetic muons ever detected with $E_{\upmu}=\unit[1]{PeV}$~\cite{IceCube:2015wro}. These data result in the lower constraint
\begin{equation}
\label{eq:constraint-c-muon-IceCube}
-4.2\times 10^{-15}<\ring{c}^{(\upmu)}\,.
\end{equation}
The latter limit is also stricter than the sensitivity that could be achieved via an estimate of synchrotron radiation losses in a future muon collider \cite{InternationalMuonCollider:2022qki}; cf.~App.~\ref{sec:sensitivity-synchrotron-radiation}.

A second non-standard process that can occur in the presence of isotropic $c$-type LV is photon decay~\cite{Liberati:2001cr,Jacobson:2001tu,Jacobson:2002hd,Jacobson:2005bg,Klinkhamer:2008ky,Hohensee:2008xz,Shao:2010wk,Altschul:2010nf,Rubtsov:2012kb,Satunin:2013an,Rubtsov:2013wwa,Diaz:2015hxa,Kalaydzhyan:2016lfo,Martinez-Huerta:2016odc,Martinez-Huerta:2017ulw,Martinez-Huerta:2017ntv,Martinez-Huerta:2017unu,Klinkhamer:2017puj,Martinez-Huerta:2019rkx,Satunin:2019gsl,Chen:2021hen,Duenkel:2021gkq,Wei:2021ite}. A photon whose energy exceeds
\begin{equation}
\label{eq:threshold-energy-photon-decay-isotropic-c}
E_{\upgamma}^{\text{th}}=\sqrt{\frac{3}{2}}\frac{m}{\sqrt{\ring{c}}}\,,
\end{equation}
decays into a fermion-antifermion pair (each of mass $m$). The form of Eq.~\eqref{eq:threshold-energy-photon-decay-isotropic-c} suggests that photon decay can occur only for positive $\ring{c}$. Note that a photon favours to decay into an $\mathrm{e^-e^+}$ pair, as the corresponding threshold energy is lower than that for a decay into a $\upmu^-\upmu^+$ pair.

The coefficient $\ring{c}$ can thus be constrained by the observation of high-energy astrophysical photons. The Large High Altitude Air Shower Observatory (LHAASO)~\cite{LHAASO:2019qwt} in China has recently detected very-high-energy photon events with energies up to $\unit[1.4]{PeV}$ \cite{Cao:2021,Li:2021tcw}, opening a new era of gamma astronomy. From this event energy, we are able to derive the following upper bounds:
\begin{equation}
\ring{c}^{(\mathrm{e})}<1.9\times 10^{-19}\,,\qquad
\ring{c}^{(\upmu)}<8.3\times 10^{-15}\,,
\end{equation}
where the second holds under the assumption $\ring{c}^{(\mathrm{e})}=0$.

\section{Combined Results}

Combining the constraints from vacuum \v{C}erenkov radiation and photon decay deduced in the last section, we can obtain two-sided limits on the isotropic $c$ coefficients:
\begin{subequations}
\begin{align}
-1.6\times 10^{-14}&<\ring{c}^{(\mathrm{e})}<1.9\times 10^{-19}\,, \\
-4.2\times 10^{-15}&<\ring{c}^{(\upmu)}<8.3\times 10^{-15}\,.
\end{align}
\end{subequations}
These numbers show that it is impossible to account for $\Delta a_{\upmu}$ through LV. Similarly, the isotropic $c$-type coefficient in electrons is already too tightly constrained by LHC searches as well as by the absence of photon decay to be able to explain the tensions of Eq.~\eqref{eq:g-2_e_val_1} (Eq.~\eqref{eq:g-2_e_val_2}) at $2\,\sigma$ ($1\,\sigma$) confidence level. The situation is illustrated in Fig.~\ref{fig:electron_muon_plot}, which confronts the anomalous magnetic moments of electrons (muons) with the threshold energies for vacuum \v{C}erenkov radiation and photon decay. The solid (dashed) black curves are determined by the threshold energy of Eq.~\eqref{eq:threshold-energy-isotropic-c} (Eq.~\eqref{eq:threshold-energy-photon-decay-isotropic-c}). The values for $\Delta a_{\mathrm{e}(\upmu)}$ to the right of the intersection points of the the solid curves and the dark blue (cyan) horizontal lines are excluded by the non-observation of vacuum \v{C}erenkov radiation. The grey regions at negative $\Delta a_{\mathrm{e}(\upmu)}$ are excluded by the absence of photon decay.

\section{Conclusions}
\label{sec:conclusions}
In this paper we explored whether the presence of a particular type of LV in the electron (muon) sector can explain the tensions between the measurements and the SM predictions for the electron (muon) anomalous magnetic moment. Focusing on spin- and direction-independent LV effects within the minimal SME, we calculated the impact of the $\ring{c}^{\ell}$ coefficients on $(g-2)_\ell$. These are constrained by the non-observation of vacuum \v{C}erenkov radiation of charged leptons and of photon decay into lepton-antilepton pairs. In particular, using LHAASO data for photons, as well as the measured maximum electron (muon) energies observed at the LHC (IceCube), we computed strict two-sided bounds on $\ring{c}^{\ell}$. As these constraints are more stringent than the ranges preferred for these coefficients by $(g-2)_\ell$, we conclude that spin-degenerate LV cannot explain the corresponding tensions.
\medskip

\end{fmffile}
{\it Acknowledgements} --- { We thank David Hertzog, Martin Hoferichter, Matthew Mewes, and Philipp Schmidt-Wellenburg for useful discussions and helpful comments. The work of A.C.~and F.K.~is supported by a Professorship Grant (PP00P2\_176884) of the Swiss National Science Foundation. M.S. is indebted to FAPEMA Universal 00830/19, CNPq Produtividade 310076/2021-8, and CAPES/Finance Code 001.}

\appendix

\section{Modified photons}
\label{sec:modified-photons}

To perform a crosscheck of the anomalous magnetic moment induced by $c$-type LV in the charged-lepton sector, Eq.~\eqref{eq:anomaly-modified-fermions}, we intend to perform an independent computation of this quantity based on LV in photons. This investigation rests upon the altered QED stated in Eq.~\eqref{eq:modified-qed} with $c^{\mu\nu}=0$. The generic field equations for photons modified by $k_F$ read~\cite{Colladay:1998fq}:
\begin{subequations}
\label{eq:field-equations-electromagnetism}
\begin{align}
M^{\alpha\delta}(p)A_{\delta}(p)&=0\,, \\[2ex]
M^{\alpha\delta}(p)&=\eta^{\alpha\delta}p^2-p^{\alpha}p^{\delta}-2(k_F)^{\alpha\beta\gamma\delta}p_{\beta}p_{\gamma}\,.
\end{align}
\end{subequations}
We focus on the set of coefficients leading to a nonbirefringent vacuum at first order in LV. The latter is parametrised by \cite{Altschul:2006zz}
\begin{equation}
\label{eq:nonbirefringent-ansatz}
(k_F)^{\alpha\beta\gamma\delta}=\frac{1}{2}(\eta^{\alpha\gamma}\tilde{k}^{\beta\delta}-\eta^{\alpha\delta}\tilde{k}^{\beta\gamma}-\eta^{\beta\gamma}\tilde{k}^{\alpha\delta}+\eta^{\beta\delta}\tilde{k}^{\alpha\gamma})\,,
\end{equation}
where the set $\tilde{k}^{\alpha\beta}\equiv (k_F)_{\gamma}^{\phantom{\gamma}\alpha\gamma\beta}$ defines a symmetric and traceless $(4\times 4)$ matrix. Inserting Eq.~\eqref{eq:nonbirefringent-ansatz} into the tensor-valued function $M^{\alpha\delta}$ of the modified field equations in Eq.~\eqref{eq:field-equations-electromagnetism} and keeping only the terms linear in $\tilde{k}^{\alpha\beta}$ leads to
\begin{subequations}
\begin{align}
\tilde{M}^{\alpha\delta}(p)A_{\delta}&=0\,, \\[1ex]
\tilde{M}^{\alpha\delta}&=(\eta^{\alpha\delta}+\tilde{k}^{\alpha\delta})(\eta^{\beta\gamma}+\tilde{k}^{\beta\gamma})q_{\beta}q_{\gamma} \notag \\
&\phantom{{}={}}-(\eta^{\alpha\beta}+\tilde{k}^{\alpha\beta})q_{\beta}(\eta^{\delta\gamma}+\tilde{k}^{\delta\gamma})q_{\gamma}+\dots\,,
\end{align}
\end{subequations}
which, by multiplying with $\delta^{\alpha}_{\phantom{\alpha}\delta}-\tilde{k}^{\alpha}_{\phantom{\alpha}\delta}/2$, can be expressed in alternative form,
\begin{subequations}
\begin{align}
0&=\tilde{A}^{\alpha}\tilde{q}_{\beta}\tilde{q}^{\beta}-\tilde{q}^{\alpha}\tilde{q}_{\gamma}\tilde{A}^{\gamma}+\dots\,, \\[1ex]
\tilde{q}^{\mu}&\equiv\left(\eta^{\mu\nu}+\frac{1}{2}\tilde{k}^{\mu\nu}\right)q_{\nu}\,,\quad \tilde{A}^{\mu}\equiv\left(\eta^{\mu\nu}+\frac{1}{2}\tilde{k}^{\mu\nu}\right)A_{\nu}\,,
\end{align}
\end{subequations}
with $\tilde{A}^{\mu}=\tilde{A}^{\mu}(\tilde{q})$ and where the ellipses indicate higher-order contributions in the LV coefficients. For convenience, we define $\tilde{F}_{\mu\nu}=\tilde{F}_{\mu\nu}(\tilde{q})\equiv-\mathrm{i}(\tilde{q}_{\mu}\tilde{A}_{\nu}-\tilde{q}_{\nu}\tilde{A}_{\mu})$ as the electromagnetic field strength tensor of the redefined four-potential in momentum space. Thus, at first order in LV, the effects of nonzero coefficients $\tilde{k}^{\alpha\beta}$ are governed by an effective metric $\tilde{\eta}^{\mu\nu}\equiv\eta^{\mu\nu}+\tilde{k}^{\mu\nu}$, since the modified photon dispersion equation reads $\tilde{q}^{\mu}\tilde{q}_{\mu}\simeq q_{\mu}\tilde{\eta}^{\mu\nu}q_{\nu}=0$.

Now, the anomalous magnetic moment caused by $\tilde{k}^{\mu\nu}$ can be inferred from the standard matrix element of Eq.~\eqref{eq:amplitude-tree-level-standard} by performing the inverse substitutions. As the Dirac fermions are unaffected by LV, the standard Gordon identity is employed. Then,
\begin{subequations}
\begin{align}
\mathrm{i}\mathcal{M}^{(0)}_{\upgamma,ss'}&=-\frac{\mathrm{i}e}{2m}\overline{u}^{(s)}\left[\tilde{P}^{\mu}(\eta_{\mu\nu}-\tilde{k}_{\mu\nu})\tilde{A}^{\nu}(q)\right. \notag \\
&\phantom{{}={}}\hspace{1.6cm}-\frac{1}{2}\left(\eta_{\mu\varrho}-\frac{1}{2}\tilde{k}_{\mu\varrho}\right)\left(\eta_{\nu\sigma}-\frac{1}{2}\tilde{k}_{\nu\sigma}\right) \notag \\
&\phantom{{}={}}\hspace{2cm}\left.\times\sigma^{\mu\nu}\tilde{F}^{\varrho\sigma}(q)\right]u^{(s')} \notag \\
&=\mathrm{i}(\mathcal{M}^{(0)}_{ss'}+\delta\mathcal{M}_{\upgamma,ss'}^{(0)})\,,
\end{align}
with the standard amplitude of Eq.~\eqref{eq:amplitude-tree-level-standard} and the alteration
\begin{align}
\delta\mathcal{M}^{(0)}_{\upgamma,ss'}&=\frac{e}{2m}\overline{u}^{(s)}\bigg[-\frac{1}{4}(\tilde{k}^{\varrho\kappa}\sigma_{\kappa}^{\phantom{\kappa}\nu}-\tilde{k}^{\nu\kappa}\sigma_{\kappa}^{\phantom{\mu}\varrho})\tilde{F}_{\varrho\nu}(q) \notag \\
&\phantom{{}={}}\hspace{1.3cm}+\tilde{k}^{\mu\nu}\tilde{P}_{\nu}\tilde{A}_{\mu}(q)\bigg]u^{(s')}\,.
\end{align}
\end{subequations}
Focusing on the part responsible for the anomalous magnetic moment, we arrive at
\begin{align}
\delta\mathcal{M}_{\upgamma,ss}^{(0)}&\supset-\frac{e}{8m}\overline{u}^{(s)}(\tilde{k}^{\varrho\kappa}\sigma_{\kappa}^{\phantom{\kappa}\nu}-\tilde{k}^{\nu\kappa}\sigma_{\kappa}^{\phantom{\mu}\varrho})u^{(s)}\tilde{F}_{\varrho\nu}(q) \notag \\
&=-\frac{e}{2m}(\tilde{k}^{11}+\tilde{k}^{22})BS^z\,,
\end{align}
such that the anomaly reads
\begin{equation}
\label{eq:anomaly-modified-photons}
a_{\upgamma}\equiv-\frac{m}{e}\frac{\delta\mathcal{M}_{\upgamma,ss}^{(0)}}{BS^z}=\frac{1}{2}(\tilde{k}^{11}+\tilde{k}^{22})\,.
\end{equation}
Note that a coordinate transformation removes $\tilde{k}^{\mu\nu}$ from the photon sector and generates $c^{\mu\nu}=-\tilde{k}^{\mu\nu}/2$ in the fermion sector \cite{Altschul:2006zz}. This correspondence between the coefficients $c^{\mu\nu}$ and $\tilde{k}^{\mu\nu}$ is evident from Eqs.~\eqref{eq:anomaly-modified-fermions}, \eqref{eq:anomaly-modified-photons}, which provides a powerful crosscheck of our results. Hence, if LV resides both in the fermion and the photon sector, the $c$ coefficients get shifted according to $c^{\mu\nu}\mapsto c^{\mu\nu}-\tilde{k}^{\mu\nu}/2$, which is why the anomaly in its complete form is given by
\begin{equation}
\label{eq:anomaly-total}
a_{\psi,\upgamma}=\left(\frac{\tilde{k}}{2}-c\right)^{11}+\left(\frac{\tilde{k}}{2}-c\right)^{22}\,.
\end{equation}

\section{Sensitivity from synchrotron radiation}
\label{sec:sensitivity-synchrotron-radiation}

A hypothetical energy loss of a charged, massive particle caused by a nonzero $\ring{c}$ may hide within the experimental uncertainty of the energy loss by synchrotron radiation. To be able to operate accelerators and storage rings, it is critical to have precise knowledge of possible synchrotron radiation losses, from which a further bound on $\ring{c}^{\ell}$ can be deduced. Reference~\cite{Hohensee:2008xz} is devoted to such an analysis for electrons based on LEP specifications.

When it comes to muons, LV could be searched for in a hypothetical $\unit[10]{TeV}$ center-of-mass muon collider~\cite{InternationalMuonCollider:2022qki}. In the following, we intend to estimate the sensitivity on $\ring{c}^{\upmu}$ based on a potential energy loss of a circulating muon via vacuum \v{C}erenkov radiation, where the latter hides in the energy loss due to synchrotron radiation. To do so, we must take additional assumptions on the specifications of the muon collider. We assume that the muon energy amounts to $\unit[5]{TeV}$ and that the bending radius of the collider is $\sim\unit[4.2]{km}$, if the LHC tunnel is to be reused. The synchrotron radiation loss of a single muon would then be $\unit[7.129]{MeV}$ per turn, which implies an energy loss per distance travelled of $\unit[5.294\times 10^{-23}]{GeV^2}$. The uncertainty on the synchrotron radiation losses at LEP was estimated to have been a fraction of $10^{-4}$ of the actual energy loss per turn~\cite{Hohensee:2008xz}. If the uncertainty in the synchrotron radiation losses of a future muon collider is comparable, the condition $\mathrm{d}W/\mathrm{d}L\leq \unit[5.294\times 10^{-27}]{GeV^2}$ based on Eq.~\eqref{eq:radiated-energy-rate} then implies a sensitivity for $\ring{c}^{(\upmu)}$ of $-1.7\times 10^{-10}$. The latter is significantly weaker than the constraint of Eq.~\eqref{eq:constraint-c-muon-IceCube}.

\bibliography{BIB}

\end{document}